\documentclass[hypertexnames=false,superscriptaddress,twocolumn,aps,prl,nobibnotes]{revtex4-2}
\usepackage{fancyref}
\usepackage{hyperref}
\usepackage{dcolumn}
\usepackage{bm}
\usepackage{tikz}
\usepackage{pgfplots}
\tikzset{>=latex}
\usepackage{amsfonts}

\usepackage{lipsum}
\usepackage{graphicx} 
\usepackage{standalone}
\usepackage[caption=false]{subfig}
\usepackage[percent]{overpic}
\usepackage{verbatim}
\usepackage{tcolorbox}
\usepackage{booktabs}
\usepackage{makecell}

\usepackage{physics} 
\usepackage{enumerate}
\usepackage{dsfont}
\usepackage{xcolor}

\usepackage[normalem]{ulem}
\usepackage[capitalise]{cleveref}

\usepackage{lipsum}

\colorlet{red}{red}

\begin{document}

\title{High-rate quantum LDPC codes for long-range-connected neutral atom registers} 

\author{Laura Pecorari}
\affiliation{University of Strasbourg and CNRS, CESQ and ISIS (UMR 7006), aQCess, 67000 Strasbourg, France}

\author{Sven Jandura}
\affiliation{University of Strasbourg and CNRS, CESQ and ISIS (UMR 7006), aQCess, 67000 Strasbourg, France}

\author{Gavin K. Brennen}
\affiliation{University of Strasbourg and CNRS, CESQ and ISIS (UMR 7006), aQCess, 67000 Strasbourg, France}
\affiliation{Center for Engineered Quantum Systems, School of Mathematical and Physical Sciences, Macquarie University, 2109 NSW, Australia}

\author{Guido Pupillo}
\email{pupillo@unistra.fr}
\affiliation{University of Strasbourg and CNRS, CESQ and ISIS (UMR 7006), aQCess, 67000 Strasbourg, France}

\date{\today}

\begin{abstract}
    High-rate quantum error correcting (QEC) codes with moderate overheads in qubit number and control complexity are highly desirable for achieving fault-tolerant quantum computing. Recently, quantum error correction has experienced significant progress both in code development and experimental realizations, with neutral atom qubit architecture rapidly establishing itself as a leading platform in the field.
    Scalable quantum computing will require processing with QEC codes that have low qubit overhead and large error suppression, and while such codes do exist, they involve a degree of non-locality that has yet to be integrated into experimental platforms. 
    In this work, we analyze a family of high-rate Low-Density Parity-Check (LDPC) codes with limited long-range interactions and outline a near-term implementation in neutral atom registers. By means of circuit-level simulations, we find that these codes outperform surface codes in all respects when the two-qubit nearest neighbour gate error probability is below $\sim 0.1\%$. 
    By using multiple laser colors, we show how these codes can be natively integrated in two-dimensional static neutral atom qubit architectures with open boundaries, where the desired long-range connectivity can be targeted via the Rydberg blockade interaction.  
\end{abstract}

\maketitle

Since Kitaev's seminal works \cite{Kitaev_2003,Dennis_2002}, the surface code has been the dominant choice for quantum error correction (QEC) as its set of check operators, or stabilizers, is simple and geometrically local enabling parallel syndrome extraction with a high tolerance to errors. However, the encoding -- only one logical qubit per code independently of the size -- is poor, posing a large resource overhead for scalable quantum computing. The surface code is but one example of a broader class of Low-Density Parity-Check (LDPC) codes \cite{Breuckmann_2021,gottesman2014faulttolerant}, other members of which retain all the good properties of the surface code such as large distance, which quantifies the number of correctable errors, while allowing for more favourable encoding rates, defined by the ratio of logical qubits to physical qubits.

\begin{figure*}[t]
    \centering
     \includegraphics[width=0.95\textwidth]{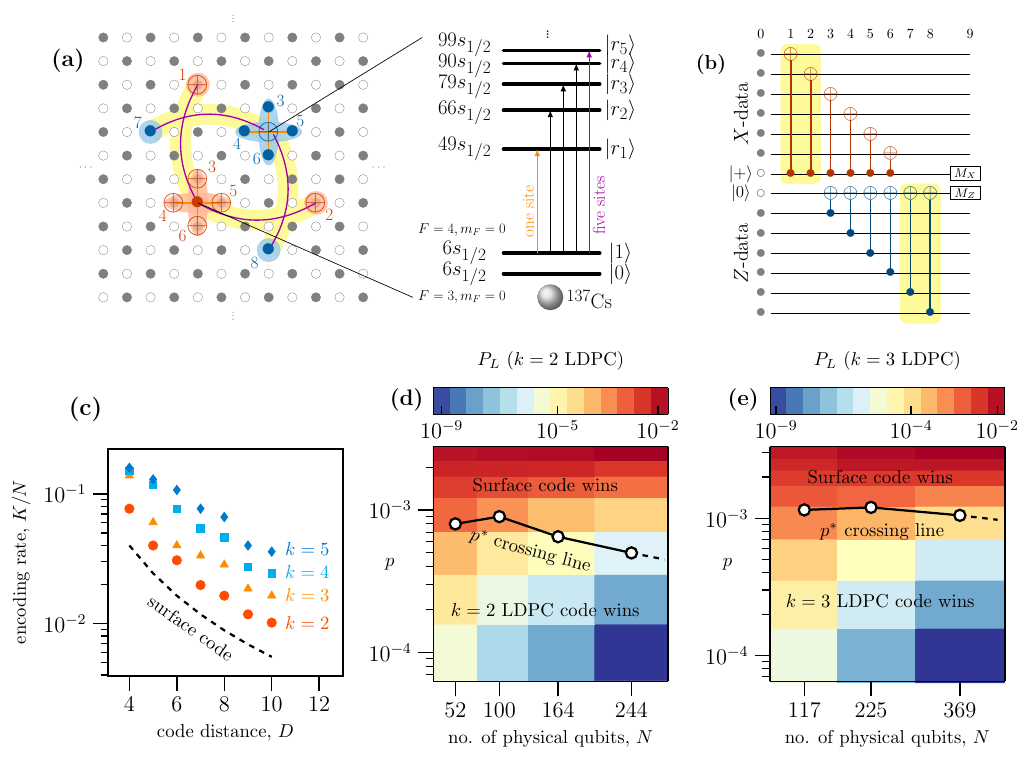}
     \caption{\textbf{Code overview for quantum error correction performance and implementation.}
     (a) Array patch with data (gray) and ancillary (white) qubits hosting one instance of the here analyzed LDPC code family. Long-range interactions are highlighted in yellow, two stabilizers are drawn (red and blue) with displayed CNOT measurement order for syndrome extraction. The inset shows energy levels and selected transitions (vertical arrows) to different Rydberg states for implementation with neutral atom qubits. Two-qubit gates between one and five lattice sites separated qubits and corresponding Rydberg transitions are coloured in orange and violet, respectively. Principal quantum numbers are chosen to minimize gate infidelity while still preserving a sufficient Rydberg blockade strength (see text). (b) Depth-$10$ syndrome measurement circuit neglecting idle errors. (c) Encoding rate, hence qubit overhead, comparison between $k$-LDPC codes and surface code (SC) given the same number of logical qubits and code distance. (d)-(e) Illustration showing logical failure probability for $k=2$- and $3$-LDPC codes as a function of nearest neighbour two-qubit gate physical error probability and number of physical qubits. Solid lines correspond to crossing probabilities below which the LDPC logical error probability gets lower than the surface code one, assuming equal number of logical and physical qubits for both codes. Dashed black line is extrapolation to larger numbers of physical qubits. }
    \label{fig:intro}
\end{figure*}

Recently, there has been intense activity benchmarking the performance of various LDPC codes both for fault-tolerant error correction \cite{Tremblay_2022, Bravyi2024, xu2023constantoverhead} and quantum computing \cite{doi:10.1126/sciadv.abn1717} with results 
converging towards the idea that “\textit{the more non-local, the better}”. However, the investigation into how to use the physics available natively in quantum computing platforms to maximally utilize the advantages of LDPC codes is still relatively at an early stage.  
Reference~\cite{Tremblay_2022} shows theoretically that a stacked two-dimensional architecture can outperform the surface code under circuit-level noise in constant-overhead hypergraph product quantum LDPC codes, provided that inter-layer cross-talks are sufficiently suppressed. Similarly, biplanar bivariate bicycle codes \cite{Bravyi2024} have recently shown high quantum error suppression with moderate long-range connectivity requirements, which make them intriguing candidates for near-term implementations on several platforms (e.g., superconducting \cite{Bravyi2024} or neutral atom \cite{poole2024architecturefastimplementationqldpc} qubits). While the recent experimental realization of both constant-rate and high-rate quantum LDPC codes via qubit shuttling with trapped ions \cite{hong2024longrangeenhanced, hong2024entanglinglogicalqubitsbreakeven} and neutral atoms ~\cite{xu2023constantoverhead,viszlai2024matching} - at the price of a considerable time overhead -- is a major advancement, the goal of identifying the \textit{best} quantum LDPC code family ultimately remains open and intrinsically hardware-dependent.

In this work, we provide a first proposal for a near term implementation in neutral atom registers of LDPC codes that is both static -- i.e. it does not require qubit shuttling -- and fast compared to existing proposals. We show that these quantum LDPC codes outperform the surface code at circuit level just using moderate non-local resources. To this purpose, we analyze a family of high-rate quantum LDPC codes built via hypergraph product (HGP) construction \cite{Tillich_2014,Kovalev_2012,Kovalev_2013,wang2022distance}. We refer to them as \textit{La-cross codes}, 
since the arrays of their stabilizer shapes, each consisting of a surface code stabilizer cross with two extra long-range interactions, is reminiscent of a long-armed cross stitch pattern (Fig.~\ref{fig:intro}(a)). The length of these interactions -- or amount of non-locality -- solely depends on the parameters of the classical seeds the quantum code is constructed from and thus is, to an extent, tunable. 
We discuss encoding capabilities and probe tolerance to errors via circuit-level simulations accounting for state preparation, measurement, single- and range-dependent two-qubit gate errors. Compared to the surface code with same number of physical and logical qubits, we obtain sub-threshold logical error probability reductions that can reach order of magnitude and increase with the number of physical qubits. 
The second part of this work is aimed at making contact with near-term experiments with neutral atom quantum registers \cite{Bloch_2008,Saffman_2010,Browaeys_2020,Henriet_2020,Morgado_2021,Bluvstein_2023}. For two-dimensional arrays with open boundaries, we show how the Rydberg blockade mechanism enables the necessary long-range gates for stabilizer measurements without need for swapping or qubit shuttling. Our error model accounts for fidelity decay as a function of gate distance, due fundamentally to the decay of the van der Waals interaction strength.
Finally, we show further improvement by adopting an error model with range-independent gate errors. In this case, the threshold increases and the onset of improvement over the surface code occurs at higher physical error probabilities. For example, the most non-local instance of La-cross codes discussed in this work shows improvement over the surface code below nearest neighbour gate error probabilities of $\sim0.5\%$, already outperforming the surface code logical failure probability by more than one order of magnitude at physical error probabilities of $\sim0.1\%$. Such a range-independent noise model may be realized by qubit shuttling in neutral atom registers \cite{xu2023constantoverhead,viszlai2024matching,hong2024longrangeenhanced}, or in different physical platforms like photonic registers with direct non-local fibre coupling \cite{bombin2021interleaving} or matter-based qubit architectures with cavity-mediated interactions \cite{jandura2023nonlocal}.

\section{Results}
\subsection{Quantum LDPC codes}
Quantum LDPC codes \cite{Breuckmann_2021} are stabilizer codes where both the number of qubits acted on by each stabilizer and the number of stabilizers acting on each qubit are constantly bounded. Stabilizers are then sparse, hence the notion of \textit{low density} invoked in their name.
The encoding rate satisfies $K/N\xrightarrow{N\rightarrow\infty}C\geq 0$, for some constant $C$, being $N$ and $K$ the number of physical and logical qubits, respectively. While constant rate, $C>0$, quantum LDPC codes exist, the better studied examples have zero rate. The surface code is an example with $K=\mathcal{O}(1)$, hence zero rate, and a code distance $D=\mathcal{O}(\sqrt{N})$. For practical use, it is not necessarily the asymptotic scaling of the rate that matters, but rather choosing a code with a favorable ratio $K/N$ for large but finite $N$, together with a large distance. 
 
In the following we review how a quantum LDPC code can be constructed via hypergraph product construction by combining two classical LDPC codes, namely classical codes with sparse checks.
Let $\mathcal{C}_i=[n_i,k_i,d_i]$ with $i=1,2$ be two classical linear codes encoding $k_i$ logical bits in $n_i$ physical bits with code distance $d_i$ (see below). Any of these codes can be represented by a matrix $H_i \in \mathbb{F}_2^{r_i \times n_i}$, called \emph{parity-check matrix}, having as many columns as physical bits and as many rows, $r_i$, as checks. Entries of $H_i$ are non-zero any time a check acts non-trivially on the corresponding bit. The number of encoded bits is then $k_i = n_i-\rank(H_i)$. The \emph{codewords} of these classical codes are vectors in the kernel of $H_i$. The minimum Hamming distance between two codewords is called the distance $d_i$. Associated with each classical code is a \emph{transposed code} $\mathcal{C}_i^T=[r_i,k_i^T,d_i^T]$ with parity-check matrix $H_i^T$.
The hypergraph product (HGP) construction \cite{Tillich_2014} combines two classical codes along with their transposed codes, $\mathcal{C}_i$ and $\mathcal{C}_i^T$, to produce a $[[N,K,D]]$ quantum stabilizer code with quantum parity-check matrix
\[
H_q=\left(\begin{array}{cc|cc}
        0 & 0 & H_1\otimes\mathbb{I}_{n_2} & \mathbb{I}_{r_1}\otimes H_2^T\\
        \mathbb{I}_{n_1}\otimes H_2 & H_1^T\otimes\mathbb{I}_{r_2} & 0 & 0
    \end{array}\right).
\]
The left block of $H_q$ describes $X$-type stabilizers, while the right one describes $Z$-type stabilizers. The total number of stabilizers then equals the number of rows of $H_q$, i.e. $n_1r_2+n_2r_1$. As before, entries of the quantum parity-check matrix are non-zero anytime a $X$ or $Z$ stabilizer acts non-trivially on the corresponding qubit. The number of physical qubits $N=n_1n_2+r_1r_2$ is half the number of columns of $H_q$, while the number of encoded logical qubits is $K=k_1k_2+k_1^Tk_2^T$. The \emph{distance} $D$ of this quantum code denotes the minimum weight of a Pauli operator commuting with all stabilizers without being itself a product of stabilizers. It can be shown that for the HGP construction the distance satisfies $D \geq \min\{d_1,d_2,d_1^T,d_2^T\}$ \cite{Tillich_2014}. The resulting quantum stabilizer code is of Calderbank-Shor-Steane (CSS) type \cite{Calderbank_1996,Steane1996}, as stabilizers are either products of only $X$ or only $Z$ Pauli operators.

For the present purposes, we have chosen the seed codes $\mathcal{C}_i$ to be cyclic codes generalizing the repetition code the surface code is built upon. Such a choice both allows for improving the encoding rate and retaining most of the intuitiveness of the surface code, at the price of a non-constant overhead in the asymptotic limit. However, this does not represent a severe issue in the prospect of implementation on near-term quantum computers, as we will discuss more quantitatively in the next sections. 

\subsection{Classical seeds and quantum layout}
\begin{figure*}[t]
    \centering
    \includegraphics[width=0.8\textwidth]{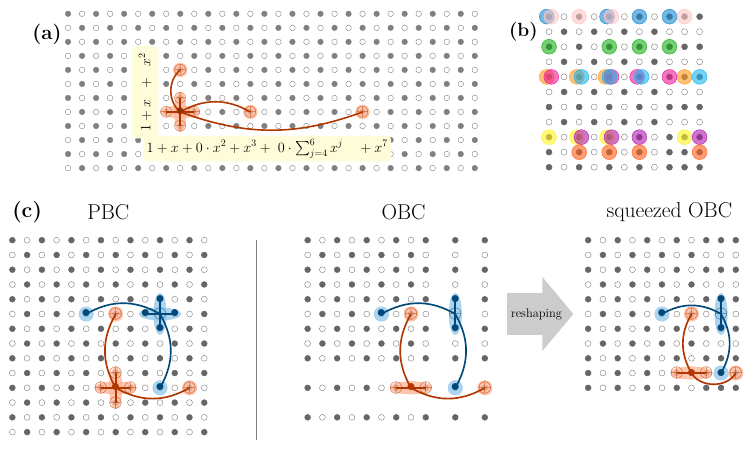}
    \caption{\textbf{Code layout, logical operators and boundary conditions.} 
    (a) Classical polynomial seeds constraining stabilizer shape and ruling one lattice direction each. Both symmetric and asymmetric configurations are allowed. The examples here refer to the cases $n_1=6,\,k_1=2$ (vertical) and $n_2=15,\,k_2=7$ (horizontal). Data(ancilla) qubits are coloured black(white). (b) The $[[65,9,4]]$ $k=3$-LDPC code with the $9$ partially overlapping horizontal logical $Z$ operators shown in different colors. (c) Array shape and two stabilizers with periodic boundary conditions (PBC), open boundary conditions (OBC) and squeezed open boundary conditions, assuming $n=7$ and $k=3$ (equal seeds). Long-range gates of boundary stabilizers are effectively shorter upon array reshaping (from length $j=5$ lattice sites to $j=4$ or $j=3$ here), which should be included in quantum error correction simulations under a range-dependent noise model.}
    \label{fig:fig2}
\end{figure*}
In this section we review how stabilizers, logical operators, and array shape of a quantum HGP code are determined from its classical seeds.
This represents a huge help in designing new LDPC codes tailored to the connectivity of the quantum hardware. 

We have chosen cyclic seed codes, i.e. codes with cyclic shift invariant codewords. A square matrix is said to be circulant if its rows are cyclic shifts of the first row \cite{Kovalev_2013}. When the parity-check matrix, $H$, is circulant, the associated code is fully specified by the first row of $H=\text{circ}(c_0,c_2,\dots,c_k,0,\dots,0)\in\mathbb{F}_2^{n\times n}$. Entries $c_i$ ($i=0,1,\dots,k$) can be mapped into coefficients of a degree-$k$ polynomial of the form $h(x)=1+\sum_{i=1}^{k}c_ix^i$. More formally, there exists a map $\mathbb{F}_2^n\rightarrow\mathbb{F}_2[x]/(x^n-1)$, being $\mathbb{F}_2[x]/(x^n-1)$ the ring of polynomials dividing $x^n-1$. This map transforms cyclic shifts in $\mathbb{F}_2^n$ into multiplications by $x$ in $\mathbb{F}_2[x]/(x^n-1)$, hence cyclic-shift-invariant codes into polynomials invariant under $x$-multiplication. Due to the ring structure, invariance under multiplication by $x$ is equivalent to invariance under multiplication of any element of the ring, a property defining the so-called \emph{ideals} of the ring. Thus, there exists a one-to-one correspondence between cyclic codes $\mathcal{C}\subseteq\mathbb{F}_2^n$ and ideals of $\mathbb{F}_2[x]/(x^n-1)$, which in turn are in one-to-one correspondence with unitary mod--$2$--divisors of $x^n-1$ having leading coefficient equal to $1$. Building blocks of length-$n$ cyclic codes then correspond to factors of $x^n-1$. For $k=1$ the repetition code is recovered. 

We note that there always exists a mapping from quantum parity-check matrix to array indexing so that the shape of the stabilizers of the code can be directly inferred from the polynomial of the classical seed, see Fig.~\ref{fig:fig2}(a). Additionally, there always exists a
basis where logical operators align along the same row or column. In contrast to the surface code where logical operators are Pauli strings stretching from boundary to boundary, for the present code family logical operators are Pauli strings – possibly with holes – which are shorter in length and larger in number [Fig.~\ref{fig:fig2}(b)]. As a second remark, we observe that, if $H$ is circulant, for any $n$ only some $k$'s are allowed, as $H$ may happen to be full-rank, and thus $k=0$. The HGP of equal seeds having circulant parity check matrix, $H\in\mathbb{F}_2^{n\times n}$, naturally leads to a quantum code with parameters $N=2n^2$, $K=2k^2$ and periodic boundary conditions. In this work we are interested in codes with open boundary conditions for the sake of experimental realization, thus we choose a full-ranked $H\in\mathbb{F}_2^{(n-k)\times n}$, which by being rectangular allows for any choice of $k$ and $n$. Quantum parameters now read $N=(n-k)^2+n^2$ and $K=k^2$ (half logical qubits get ``lost"), with consequent array shape shown in Fig.~\ref{fig:fig2}(c), which can be squeezed to restore a full square configuration. A consequent effect is that squeezed stabilizers have effectively shorter \emph{legs} [see Fig.~\ref{fig:fig2}(c) right panel], where a leg is defined as the distance between the central ancilla and a non-local data qubit of the stabilizer. Depending on the exact position of the boundary stabilizer, this effective shortening can even result in legs of different length in different directions [see, e.g., the red stabilizer in Fig.~\ref{fig:fig2}(c) right panel]. Interestingly, this effective leg shortening improves the circuit-level error correction performance (see below) and is later accounted for in quantum error correction simulations. 
More details regarding seeds choice and code construction are provided in Appendix A.

In the following we stick to a sub-family of HGP codes with equal seed polynomials  of the form $h(x)=1+x+x^k$ and consequently weight-$6$ stabilizers which we study for different values of $k$. The $k=2$ instance of this family has recently been studied in \cite{hong2024longrangeenhanced}, where the possibility of implementation via qubit shuttling is discussed. This polynomial choice provides quantum codes with high rate, low stabilizer weight and moderate non-locality both in terms of \textit{range} and \textit{number} of long-range interactions, while allowing for a similar implementation scheme as the surface code.  We mention in passing that the similar polynomial $\bar{h}(x)=1+x^k$ leads to codes with a shorter distance than $h(x)=1+x+x^k$ and shows no apparent improvement in the overhead over the surface code. For example, the HGP of two $[9,3,3]$ classical codes, i.e. $1+x^3$, leads to a quantum code with $N=162$ (periodic boundary conditions) or $N=117$ (open boundary conditions). These code parameters exactly match the number of physical qubits of $K$ copies of surface codes, thus not allowing for any overhead saving, independently of the choice of boundary conditions.

\subsection{Error models}
\begin{figure*}[t]
    \centering
    \includegraphics[width=\textwidth]{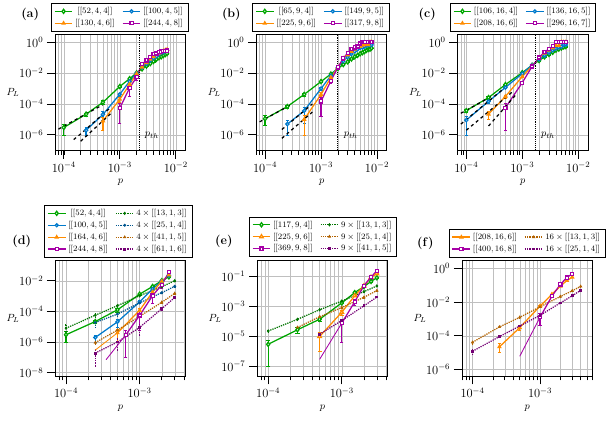}
        \caption{
        \textbf{Error correction performance under hardware-specific noise model.} 
        Cumulative logical error probability normalized by the number, $D$, of rounds $P_L(p)=(1-p_L(p))^{1/D}$ for $k=2$ (a), $k=3$ (b), $k=4$ (c) vs nearest-neighbor two-qubit gate error probability $p$, under circuit-level depolarizing errors with hardware-specific noise. Error bars correspond to standard deviations $\sigma_{P_L}=\sqrt{P_L(1-P_L)/(\text{shots})}$. Vertical dotted lines indicate the approximate location of the threshold probability $p_{th}$. Dashed black lines are added to show the qualitative good agreement of the decoding curves with the expected asymptotic scaling. (d)-(f) Logical error probability comparison against the surface code given same number of logical and physical qubits. For larger-distance codes extrapolations to lower logical error probabilities are shown (solid lines) to guide the eye. Surface code simulations have been performed under depolarizing noise with same single- and nearest neighbour two-qubit gate and SPAM errors as the LDPC codes.}
    \label{fig:decoding}
\end{figure*}

We perform quantum error correction with data and ancilla qubits placed in the same square array, analogous to an unrotated surface code with open boundaries. To measure a stabilizer, a CZ gate has to be applied between the ancilla qubit, located in the middle of the cross describing the stabilizers, and the six data qubits [see Fig.~\ref{fig:intro}(a/b)]. For the four data qubits which directly neighbor the ancilla qubits, this can be done simultaneously for all stabilizers \cite{Tomita_2014}, while the long-range CZ gates to the remaining data qubits have to be applied separately for $X$ and $Z$ stabilizers, leading to a stabilizer measurement circuit of depth 10 [see Fig.~\ref{fig:intro}(b)]. Such gate ordering for the syndrome extraction circuit also ensures robustness against hook errors when open boundary conditions are enforced (see Appendix B).

The code tolerance against errors is probed via numerical simulations under circuit-level depolarizing noise, which is chosen to directly compare with existing literature \cite{Bravyi2024,xu2023constantoverhead}. 
We assume uniformly distributed Pauli errors drawn from $\{X,Z,Y\}$ and $\{I,X,Z,Y\}^{\otimes2}\backslash \{I\otimes I\}$ with probability $p_1/3$ and $p_2/15$ for single and two-qubit errors, respectively. We simulate $D$ rounds of syndrome measurements, using Stim \cite{gidney2021stim} to sample from the code circuit and Belief Propagation with Ordered Statistics Decoder (BP+OSD) \cite{Roffe_2020,Roffe_LDPC_Python_tools_2022} to process syndrome information (see Methods). The latter is commonly regarded as one of the most viable approaches to decode arbitrary quantum LDPC codes.

For the error model above, we denote our error probabilities as $\{p_1,p_2(j),p_p,p_m\}$ where $p_1$ is the aforementioned single-qubit error probability, $p_2(j)$ is the two-qubit error probability for separations of $j$ in units of lattice spacing between control and target, $p_p$ and $p_m$ are the preparation and measurement error probabilities respectively. Idle errors are neglected. We focus most of our analysis on noise ratios appropriate for neutral atom hardware based on numbers reported in experiments for state preparation and measurement (SPAM) errors \cite{Graham_2022}, and gate errors \cite{Evered_2023, Graham_2022}, by adopting the following \emph{hardware-specific noise} parameters $\{p=p_2(1),p_2(j)=c_j p,p_1=p/10,p_p=p_m=2p\}$, with $c_j$ the proportionality constants between long-range two-qubit gate error probabilities and the nearest neighbour two-qubit gate error probability (see below). The choice of linear dependence is justified in the Methods section. Experimentally, atom arrays can be directly loaded in the reshaped configuration discussed before for squeezed open boundary conditions [Fig.~\ref{fig:fig2}(c)]. Consequently, under the range-dependent noise model, boundary long-range gates get shorter and the associated fidelities improve at the boundaries. Such a favorable finite-size effect is particularly relevant for small atom arrays.
Subsequently, anticipating potential implementation with other platforms and mostly for comparison with existing literature, we provide numerical results for an \emph{hardware-agnostic noise} parameter set using $\{p\equiv p_1=p_2(j)\forall j, p_p=p_m=0\}$, i.e. where all unitary gate errors are treated as equal (Cfr. Ref.\cite{Fowler_2012}). Measurement and reset gates have hardware-dependent fidelities, hence we just set the SPAM error strength to zero.

\subsection{Performance of La-cross codes}
Overhead reduction against the surface code is the main motivation for adopting quantum LDPC codes. As shown in Fig.~\ref{fig:intro}(c), the present code family offers a significant advantage in terms of encoding rate, and hence qubit overhead, when compared to surface codes having the same number of logical qubits and distance, although asymptotically both still scale like $K=\mathcal{O}(1),\,D=\mathcal{O}(\sqrt{N})$. Along with overhead saving, the present code family also shows advantage over the surface code given equal number of physical and logical qubits, offering larger code distance and lower logical error probability for sufficiently small physical error probabilities, as shown below. 

In Fig.~\ref{fig:decoding}(a)-(c) we first present the error correction performance for several different sizes of La-cross codes under hardware-specific noise and defer the hardware-agnostic case to the Methods section (Fig.~\ref{fig:Adecoding}(a)-(c)). We plot the {\it cumulative} logical error probability, i.e. the probability that any of the $K$ logical qubits fails, normalized by the number of rounds, $P_L=1-(1-p_L)^{1/D}$ vs the nearest-neighbor two-qubit gate error probability $p$, with $p_L=\text{errors}/\text{shots}$, for $k=2,3,4$-LDPC codes. This is consistent with real experiments where one wants \textit{all} logical qubits to be protected at the same time. We find nearest-neighbor two-qubit gate error probability thresholds $p_{th}^{k=2}\approx0.22\%$, $p_{th}^{k=3}\approx0.20\%$, $p_{th}^{k=4}\approx0.17\%$, which for simulations with hardware-agnostic noise further improve to $p_{th}^{k=2}\approx0.38\%$, $p_{th}^{k=3}\approx0.45\%$, $p_{th}^{k=4}\approx0.5\%$ (see Methods). While in the latter case $p_{th}^{k}$ increases with the 
degree of non-locality $k$, in the hardware-specific case long-range gates get longer and thus more faulty (see below) and so $p_{th}^{k}$ decreases with $k$. 
In all cases, the slope of the decoding curves is found to be consistent within good agreement (black dashed lines) with the expected behaviour in the deep sub-threshold regime: $P_L(p)\approx A\left(p/p_{th}\right)^{D_e}$, being $D_e=\left\lfloor\frac{D+1}{2} \right\rfloor$ the effective distance of the code, namely the length of the minimal physical error chain triggering a logical error. 

We show in Fig.~\ref{fig:decoding}(d)-(f) comparisons of error correction performance for the $k=2,\,3,\,4$-LDPC codes with the surface code with equal number of logical and physical qubits. We find that in all cases a crossing occurs between the LDPC and surface code decoding curves at a given nearest-neighbor gate error probability $p^*\sim10^{-3}$, with the LDPC achieving lower logical errors for $p<p^*$ [see also Fig.~\ref{fig:intro}(d)-(e)]. 
The crossing value
$p^*\sim10^{-3}$ is already within experimental reach, despite the penalty on long-range gates we have enforced. 

We observe that $p^*$ slowly decreases as the distance increases. In the sub-threshold regime, the logical error probability for both codes scales as $P_L(p)\approx A\left(p/p_{th}\right)^{ D/2}=A\left(p/p_{th}\right)^{\beta\sqrt{N}}$,
with $\beta=\mathcal{O}(1)$ and $A$ is the logical error probability extrapolated from the power-law behaviour up to the threshold. Defining the two $k$ dependent values $\beta_{\text{sc}}=(2\sqrt{2K})^{-1}$ and $\beta_{\text{ldpc}}$, with the former determined by  partitioning the $N$ physical qubits into $K$ surface codes so the number of logical qubits are equal for both, a simple argument (see Methods) shows that
\[
    p^*=\left(\frac{(p_{th}^{\text{sc}})^{\beta_{\text{sc}}}}{(p_{th}^{\text{ldpc}})^{\beta_{\text{ldpc}}}}\right)^{\frac{1}{\beta_{\text{sc}}-\beta_{\text{ldpc}}}}\left[\left(\frac{A_{\text{ldpc}}}{A_{\text{sc}}}\right)^{\frac{1}{\beta_{\text{sc}}-\beta_{\text{ldpc}}}}\right]^{\frac{1}{\sqrt{N}}}.
\]
Therefore, since $\beta_{\text{sc}}<\beta_{\text{ldpc}}$ by construction and from data extrapolation $0<(A_{\text{ldpc}}/A_{\text{sc}})<1$, we find that $p^*$ decreases as the size $N$ increases, but converges to a constant greater than zero asymptotically. This ensures that the LDPC codes {\emph {always}} offer lower logical error probability for sufficiently small physical error probabilities with respect to the surface code, given the same number of logical and physical qubits.

In Appendix C we compare the La-cross codes to other quantum LDPC codes in terms of qubit overhead, type of connectivity and geometrical layout.

\subsection{Implementation with neutral atom qubits}
\begin{figure}
    \centering
    \includegraphics{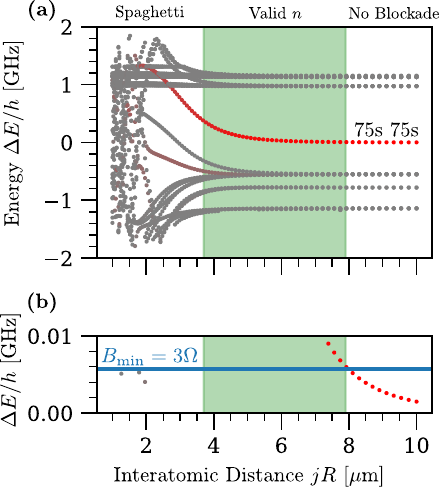}
    \caption{
    \textbf{Two-atom Hamiltonian eigenenergies.} 
    Pair state interaction energy $\Delta E$ calculated as a function of the interatomic distance $jR$ for different eigenstates of the two-atom Hamiltonian. Red points are eigenenergies of eigenstates with the largest overlap with the $\ket{75s, 75s}$ state. Panel (b) shows zoom-in of panel (a) on the range $0 \leq \Delta E/h \leq 10$ MHz. Blue line shows the minimal interaction $B=3\Omega$ which we assume to be necessary for a Rydberg blockade gate (We take the realistic value $\Omega = 2\pi \times 1.9$ MHz \cite{Saffman_2010}) The range of valid distances in which there is a sufficient blockade ($B \geq 3\Omega$) but we have not yet entered the ``Spaghetti'' regime, where the perturbative treatment of the dipole-dipole interaction between the atoms breaks down (see Methods), is shaded in green.}
    \label{fig:molecular_potentials}
\end{figure}

In the following we discuss the implementation of our family of La-cross LDPC codes on neutral atom quantum computers. In particular, we show how long-range connectivity can be realized natively via Rydberg-blockade interactions involving highly-excited electronic Rydberg states. We rely on the standard blockade gate \cite{PhysRevLett.85.2208}, consisting of one laser-excited atom shifting the Rydberg states of a neighboring atom off resonance via strong Rydberg-Rydberg interaction (\textit{Rydberg blockade} mechanism) to accumulate the desired two-qubit gate phase shift. In fact, while several other gate protocols exist \cite{PhysRevA.89.030301,Petrosyan_2017,PhysRevLett.123.170503,he2021multiplequbit}, such a gate scheme currently represents the most robust approach for performing high-fidelity two-qubit gates with neutral atom qubits \cite{Bluvstein_2023,Ma_2023,Scholl_2023,radnaev2024universalneutralatomquantumcomputer,Cao_2024}.

For concreteness, we here assume an array of $^{137}$Cs atoms with lattice spacing $R$ and take $\ket{0} = \ket{6s_{1/2},\,F=3,\,m_F=0}$ and $\ket{1} = \ket{6s_{1/2},\,F=4,\,m_F=0}$. To perform a CZ gate between two atoms with a separation of $j$ lattice sites we couple the state $\ket{1}$ to a Rydberg state $\ket{r_j} = \ket{n_js}$ using a two photon transition via the intermediate $\ket{7p_{1/2}}$ state with effective Rabi frequency $\Omega_j$. We choose the laser phase $\varphi(t)$ according to the time-optimal (TO) protocol \cite{Jandura_2022,PhysRevResearch.4.033019} in order to minimize gate duration and Rydberg scattering. Note that we propose to use \emph{different} principal quantum numbers $n_j$ for different interatomic distances. Hereafter in this section the symbol $n$ will refer to principal quantum numbers and not to classical code parameters. 

The atoms interact via a dipole-dipole interaction, which for large atomic distances $jR$ can be perturbatively treated as a van der Waals interaction $\sum_j B_j \ket{r_jr_j}\bra{r_jr_j}$, where the \textit{blockade} strength $B_j$ scales as $B_j \propto n_j^{11}/(jR)^6$. There are two constraints on our choice of $n_j$: First, $n_j$ must be large enough to ensure a sufficient Rydberg blockade. While the TO protocol was originally designed in the limit $B_j\gg \Omega_j$, small variations of the phase profile $\varphi(t)$ allow to implement a CZ gate as long as $B_j \gtrsim \Omega_j$. For concreteness, we here require $B_j \geq 3\Omega_j$, striking a compromise between allowing finite interaction strengths $B_j$ while preserving the qualitative behavior of the $B_j \gg \Omega_j$ limit. The second constraint is that $n_j$ must be small enough such that the perturbative treatment of the dipole-dipole interaction is valid. For too large $n_j$ we enter the so-called ``Spaghetti'' regime, in which the eigenstates of the two atom Hamiltonian cannot be approximated by product states anymore. Both the lower and the upper bound of $n_j$ depend on the interatomic distance $jR$. This is exemplified in Fig.~\ref{fig:molecular_potentials}, which shows the eigenenergies of the two-atom Hamiltonian near the $\ket{75s,75s}$ state \cite{_ibali__2017}. The shaded green area marks the range of interatomic distances $jR$ for which $n_j=75$ allows for a sufficient Rydberg blockade without entering the ``Spaghetti'' regime.

For a realistic analysis of the logical error probability it is crucial to understand how the physical error probability scales with the number of lattice sites $j$ between the atoms. To estimate this, we assume that the only two sources of infidelity are the decay of the Rydberg states with decay rates $\gamma_j$ and dephasing of the Rydberg states due to a Doppler shift arising from thermal motion of the atoms at $T=10$ $\mu$K. The infidelity of a time-optimal CZ gate is then given by \cite{PRXQuantum.4.020336}
\begin{equation}
    1-F_j = 2.96\frac{\gamma_j}{\Omega_j} + 7.12 \frac{\Delta_{\mathrm{Doppler}}^2}{\Omega_j^2}
    \label{eq:infidelity}
\end{equation}
where $\Delta_{\mathrm{Doppler}} = k_{\mathrm{eff}} \sqrt{k_B T/m}$ and $k_\mathrm{eff}$ is the effective wave vector of the two photon transition. To find the value $n_j$ with the lowest infidelity $1-F_j$ we assume realistic values of $R=3$ $\mu$m and $1/\gamma = 430$ $\mu$s at $n=75$ and use the scalings $\Omega \propto n^{-3/2}$ and $\gamma\propto n^{-3}$ \cite{TFGallagher_1988}. For each $j$ we now consider different values of the laser intensity (characterized by the value of $\Omega$ at a reference principal quantum number $n=75$) and numerically minimize $1-F_j$ over $n_j$, constraining $n_j$ to be large enough to achieve a sufficient blockade and small enough to not enter the 
``Spaghetti'' regime (see Methods). We find that for each $j$ the relationship $1-F_j \approx c_j (1-F_1)$ holds, with the constant $c_j$ increasing with $j$. This scaling forms the basis of our numerical simulation of the logical error probability.

We note that also the gate duration $\tau$ increases with increasing interatomic distance $j$, scaling as $\tau \propto (jR)^{18/25}$ (see Methods). With demonstrated gate durations around $250$ ns for nearest neighbor atoms \cite{Evered_2023}, a gate between atoms separated by $j=7$ lattice sites (the maximum considered above), could be implemented in approximately $1$ $\mu$s.  

The implementation on Rydberg atoms dictates that different two-qubit gates can only be implemented in parallel when the qubit pairs are sufficiently separated, i.e. outside a \textit{blockade sphere} of radius $R_b\propto\Omega^{-1/6}\gtrsim2R$, since otherwise an atom in the Rydberg state in one qubit pair might blockade an atom in another qubit pair \cite{9499945}. A safe option is to divide the lattice into a rectangular grid of square subregions each of size $2(k+1)\times 2(k+1)$ and measure each subregion stabilizers, one by one, in parallel over the grid. This will enable measuring all the stabilizers in $4(k+1)^2$ measurement rounds.

Finally, we mention that while the values of the $c_j$ are specific for Cs atoms at $10$ $\mu$K, we expect qualitatively similar behavior for other atoms at other temperatures. Dephasing of the Rydberg state arising from sources other than Doppler shifts and optical pumping by black body radiation could be easily incorporated in our model by adding an additional term in Eq.~\eqref{eq:infidelity}. Additional dephasing errors would reduce the optimal principal quantum numbers $n_j$, while additional decay due to black body radiation would increase them. Other error sources, such as leakage to other states, would require a more elaborate error model. However, most error sources become more detrimental with longer gate durations, and thus with larger interatomic distances. This is in accordance with our assumption that gate infidelities increase with $j$.

\section{Discussion}
We have developed an integrated approach to neutral atom QEC exploiting flexible data/ancilla qubit layout together with tunable long-range gates using multiple laser colors addressing distinct Rydberg states.
Our analysis focuses on a sub-family of HGP quantum LDPC codes with high rate, low stabilizer weight, and moderate non-locality. All these features make them promising candidates to be implemented on near-term neutral atom quantum computers with all qubits in place. 
The limiting gate time is set by the slowest elementary gate, which is typically state preparation and measurement. Since all ancilla measurements can be deferred to the end of a round of stabilizers circuits, the overall time to perform a round of stabilizer measurements for La-cross codes, even using restricted parallelization, is notably shorter than using qubit shuttling. For example, for the largest code discussed in this work, $[[400,16,8]]$, we conservatively estimate the single round time to be $0.6$ ms, which is approximately one order of magnitude shorter than the single round movement time costs estimated in Ref.~\cite{viszlai2024matching}.

We have examined the performance of these codes via numerical simulations under circuit-level depolarizing noise, which is a useful tool for benchmarking the code performance and making comparisons against the surface code. Noise models beyond depolarizing noise, appearing in many realistic situations, are expected to improve our results. For example, in Rydberg atom arrays it has been predicted that up to $98\%$ of errors can be converted to erasure errors \cite{Wu_2022,Ma_2023,Scholl_2023}, implying much better scaling of the logical error rate with physical probability. Other noise biases \cite{Roffe_2023,Dua_2024} can easily be incorporated into our construction in full analogy with the XZZX surface code \cite{Bonilla_Ataides_2021}. We defer the problem of addressing these issues to future work. 

A severe bottleneck for QEC experiments can be fast real-time decoding of syndrome information. In this work we have used a state-of-the-art decoding technique at the time of submission, namely a BP+OSD decoder, which can be flexibly applied to arbitrary quantum LDPC codes at the price of a severe time overhead in the OSD part. Recently, new decoders have been proposed to tackle the problem of fast and reliable syndrome information processing in quantum LDPC codes, such as Ambiguity Clustering \cite{Wolanski:2024bme}, Belief Propagation with Guided Decimation Guessing (BP+GDG) \cite{Gong:2024elt}, Belief Propagation with Localized Statistics Decoding (BP+LSD) \cite{Hillmann:2024gmk} and Belief Propagation with Ordered Tanner Forest (BP+OTF) \cite{iOlius:2024xos}. These decoders are faster and therefore may ultimately be preferred over BP+OSD. 

Finally, while we have focused here on how to achieve long-lived quantum memory, quantum computation will require implementing fault-tolerant logical gates. Various proposals have been made to perform logical gates with LDPC codes including using non-destructive Pauli measurements assisted with non-local gates \cite{doi:10.1126/sciadv.abn1717} and code switching to another stabilizer code like the surface code \cite{xu2023constantoverhead}. Recent work shows a way to perform transversally logical Hadamard (H) and logical CZ gates within the same array patch of HGP codes \cite{Quintavalle2023partitioningqubits}. We note that La-cross codes with open boundaries are square but not symmetric HGP codes and they may allow for transversal H and CZ gates by trapping the atoms in a folded triangular configuration in a static plane without increasing the connectivity requirements. Alternatively, one can use our QEC method for static memory and then physically transport qubits for performing logical computation, as demonstrated in Refs.~\cite{Bluvstein_2023} for transversal CNOT gates between different array patches of surface codes.

\section{Acknowledgements}
We gratefully acknowledge discussions with Shannon Whitlock. This research has received funding from the European Union’s Horizon 2020 research and innovation programme under the Marie Sk{\l}odowska-Curie project 955479 (MOQS), the Horizon Europe programme HORIZON-CL4-2021-DIGITAL-EMERGING-01-30 via the project 101070144 (EuRyQa) and from the French National Research Agency under the Investments of the Future Program projects ANR-21-ESRE-0032 (aQCess), ANR-22-CE47-0013-02 (CLIMAQS), and ANR-22-CMAS-0001 France 2030 (QuanTEdu-France). G.K.B. acknowledges support from the Australian Research Council Centre of Excellence for Engineered Quantum Systems (Grant No. CE 170100009). Computing time was provided by the High-Performance Computing Center of the University of Strasbourg. Part of the computing resources were funded by the Equipex Equip@Meso project (Programme Investissements d'Avenir) and the CPER Alsacalcul/Big Data.

\emph{After submitting this manuscript to the arXiv, two related proposals addressing the issue of two-dimensional static implementation of bivariate bycicle codes with neutral atoms appeared on the arXiv, exploiting less frequent measurements of longest-range stabilizers \cite{Berthusen:2024hit} and array folding \cite{poole2024architecturefastimplementationqldpc}.}

\bibliography{references}

\section{Methods}
\subsection{Quantum error correction simulations}
In the following we provide further detail about quantum error correction simulations. The qubit register is firstly initialized to $|0\rangle$, Hadamard gates are acted on $X$-type ancilla qubits and subsequent CNOT gates are applied according to the order prescription specified in the main text (Fig.~\ref{fig:intro}(a),(b)). We simulate as many rounds of syndrome measurements as the code distance, with ancilla measurement and reset after any round and data measurement occurring only after the last round. State preparation and measurement bit-flip errors and single- and two-qubit gate depolarizing errors are applied with probabilities specified in the main text, under both hardware-specific and agnostic noise model, which we show here in Fig.~\ref{fig:Adecoding}. Idle errors are always neglected.

Belief Propagation with Ordered Statistics Decoder (BP+OSD) \cite{Roffe_2020,Roffe_LDPC_Python_tools_2022} has been used to obtain all the decoding plots shown in the main text. We have optimized over the decoder parameters and opted for the minimum-sum variant of Belief Propagation with a scaling factor of $s=1.0$ for all LDPC codes. The number of iterations of Belief Propagation was found to be almost irrelevant, so we have fixed it to $4$. Ordered Statistics Decoding was performed in combination sweep mode up to order $1$ to speed up the decoding process, upon testing up to order $10$ without finding any relevant improvement. Monte Carlo samplings have been performed using the Sinter library with $10^3$ decoder maximal number of errors cutoff and increasing maximal number of samples cutoff with decreasing physical error rate, $\sim10^4-10^7$, compatibly with the system sizes.

Surface code simulations are performed under the same noise model of the LDPC codes they are compared to, performing $D_{\text{ldpc}}$ rounds of stabilizer measurements so to keep qubits alive for the same amount of time for both codes. Logical error probability is computed as $P_L(p)=1-(1-P_L^1(p))^K$, being $P_L^1(p)$ the single surface code logical error probability. For the fairest comparison, we want to compare our (LDPC code, BP+OSD decoder)-pair against the surface code with its best decoder. Given the impractically huge time overhead of the maximum-likelihood decoder, we have opted for decoding the surface code with the BP+OSD decoder with optimized scaling factor $s=0.625$, which was found to outperform the minimum-weight perfect matching decoder with open boundaries. Thus, with safe confidence, we claim that our comparing argument can only benefit from better decoders for the LDPC codes.

\subsection{Analytical sub-threshold error estimation}
The crossing point between LDPC and surface code decoding curves, $p^*$, given equal number of logical and physical qubits decreases as the distance increases. We can provide a quantitative and decoder-independent estimate of this behaviour via the following analytical argument. In the sub-threshold regime, the logical error probability scales as  
\begin{equation}
    \label{eq:pl_scaling}
    P_L(p)\approx A\left(\frac{p}{p_{th}}\right)^{\alpha D}=A\left(\frac{p}{p_{th}}\right)^{\beta\sqrt{N}}
\end{equation}
with $\alpha=1/2$ asymptotically and $\beta=\mathcal{O}(1)$ being code dependent. For either code, the distance as a function of $N$ can be computed from
\begin{align}
    \label{eq:Ns}
    \begin{split}
    N&=(\lambda_k D_{\text{ldpc}}-k)^2+(\lambda_k D_{\text{ldpc}})^2\approx2\lambda_k^2D_{\text{ldpc}}^2 \\
    N&=K((D_{\text{sc}}-1)^2+D_{\text{sc}}^2)\approx2KD_{\text{sc}}^2 
    \end{split}
\end{align}
being $\lambda_k$ some constant depending on the order $k$ of the LDPC code, for which -- we recall -- the code distance is $D=\mathcal{O}(\sqrt{N})$ and $K$ constant for a given code.
The crossing rate can be computed by requiring
\begin{align}
    \label{eq:pl_equal}
    \begin{split}
        A_{\text{sc}}\left(\frac{p^*}{p_{th}^{\text{sc}}}\right)^{\beta_{\text{sc}}\sqrt{N}}\overset{!}{=}A_{\text{ldpc}}\left(\frac{p^*}{p_{th}^{\text{ldpc}}}\right)^{\beta_{\text{ldpc}}\sqrt{N}},    
    \end{split}
\end{align}
leading to
\begin{equation}
\label{eq:p_cross}
    p^*=\left(\frac{(p_{th}^{\text{sc}})^{\beta_{\text{sc}}}}{(p_{th}^{\text{ldpc}})^{\beta_{\text{ldpc}}}}\right)^{\frac{1}{\beta_{\text{sc}}-\beta_{\text{ldpc}}}}\left[\left(\frac{A_{\text{ldpc}}}{A_{\text{sc}}}\right)^{\frac{1}{\beta_{\text{sc}}-\beta_{\text{ldpc}}}}\right]^{\frac{1}{\sqrt{N}}}.
\end{equation}
Therefore, when $0<(A_{\text{ldpc}}/(A_{\text{sc}}))^{1/(\beta_{\text{sc}}-\beta_{\text{ldpc}})}<1$ the crossing point decreases with $N$ for $\beta_{\text{sc}}<\beta_{\text{ldpc}}$, which is always our case by construction.

\subsection{Realistic Infidelity Calculations}
In the following we detail our estimate of the gate error for an implementation on Rydberg atoms. For a given interatomic distance $jR$ and a given laser intensity, characterized by the Rabi frequency $\Omega$ at a fixed reference $n$, e.g. $n=75$, we proceed as follows: We first use the AtomicRydbergCalculator (ARC) \cite{_ibali__2017} to determine the lowest $n_j$ for which we still obtain $B_j \geq 3\Omega_j$ (note that both $B_j$ and $\Omega_j$ change with $n_j$), as well as the highest $n_j$ which is admissible without entering the Spaghetti regime. We define the start of the Spaghetti regime as the smallest value of $n_j$ for which there is an eigenstate of the two-atom Hamiltonian which has a smaller pair state interaction energy $|\Delta E|$ then the perturbed $\ket{75s,75s}$ state We only consider states with a non-zero dipole-dipole coupling to the intermediate $\ket{7p_{1/2}}$ state in this comparison.  

\begin{figure}[htb]
    \centering
    \includegraphics{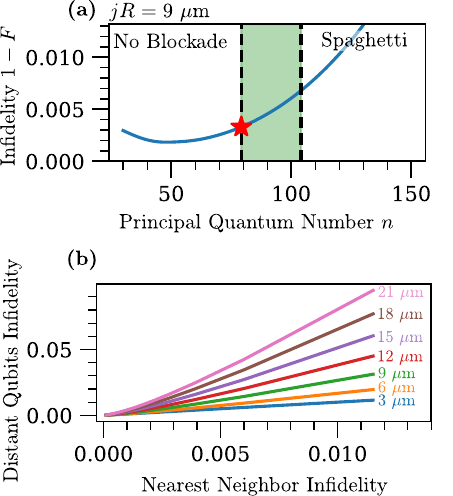}
    \caption{
    \textbf{Long-range gates modelling.} 
    (a) Infidelity for a CZ gate between two Cs atoms at $jR=9$ $\mu$m as a function of the principle quantum number $n_j$ (blue line). The shaded green area shows the allowed values if $n_j$ in which there is a sufficient Rydberg blockade but we have not yet entered the Spaghetti regime. Red start shows the minimal infidelity for valid $n_j$, which is achieved at the lowest possible $n_j$ allowing for a sufficient Rydberg blockade.
    (b) The infidelity of a CZ gate at interatomic distances between 3 $\mu$m  and $21 \mu$m ($1 \leq j \leq 7$) as a function of the nearest neighbor infidelity (distance 3 $\mu$m).}
    \label{fig:rydberg_gate_errors}
\end{figure}

\begin{figure*}[ht]
    \centering
    \includegraphics[width=\textwidth]{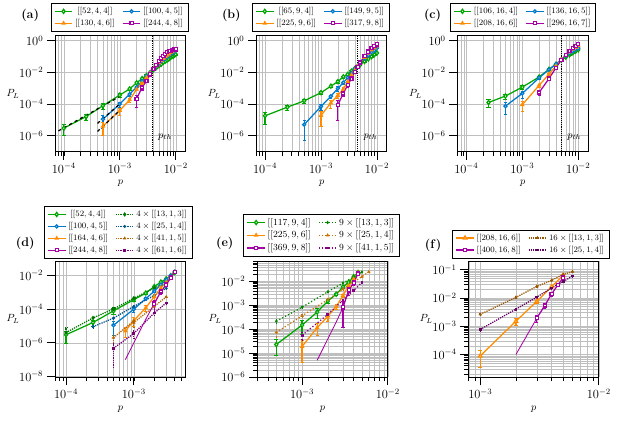}
    \caption{
    \textbf{Error correction performance under hardware-agnostic noise model.} 
    Cumulative logical error probability normalized by the number, $D$, of rounds $P_L(p)=(1-p_L(p))^{1/D}$ for $k=2$ (a), $k=3$ (b), $k=4$ (c) under circuit-level depolarizing errors with hardware-agnostic noise. Error bars correspond to standard deviations $\sigma_{P_L}=\sqrt{P_L(1-P_L)/(\text{shots})}$. Vertical dotted lines indicate the approximate location of the threshold probability $p_{th}$. For $k=2$ only, dashed black lines are added to show the qualitative good agreement of the decoding curves with the expected asymptotic scaling. (d)-(f) Logical error probability comparison against the surface code given same number of logical and physical qubits. Surface code simulations have been performed under depolarizing noise with the same error model of LDPC codes.}
    \label{fig:Adecoding}
\end{figure*}

Having established the upper and the lower bound of $n_j$, we then minimize the infidelity $1-F_j$ [Eq.~\eqref{eq:infidelity}] over this range. Fig.~\ref{fig:rydberg_gate_errors}(a) shows an example of this for an interatomic distance of $jR=9$ $\mu$m with a laser intensity such that at $n=75$ we have $\Omega=2\pi \times 1.9$ MHz \cite{PhysRevLett.123.230501}. In this example, as in all cases considered by us, the lowest infidelity is achieved at the lower bound of the allowed values of $n$. 

In Fig.~\ref{fig:rydberg_gate_errors} we vary the Rabi frequency $\Omega$ at $n=75$ from $2\pi \times 0.5$ MHz to $2\pi \times 15$ MHz and  compare the gate error $1-F_j$ for gates over $j$ lattice sites as a function of gate error $1-F_1$ for nearest neighbor gates. The ratios $c_j = (1-F_j)/(1-F_1)$ are approximately independent of $\Omega$ and given by $c_1 = 1$, $c_2 \approx 1.6$, $c_3\approx 2.5$, $c_4\approx 3.6$, $c_5\approx 4.8$, $c_6\approx 6.1$, $c_7\approx 7.5$.

Since the lowest fidelity is always achieved by the lowest value of $n$ which still gives a sufficient blockade, the ratio $B/\Omega \propto n^{25/2}/(jR)^6$ is constant, so that the optimal $n$ scales as $n_j \propto (jR)^{12/25}$. The pulse duration then scales as $\tau \propto \Omega^{-1} \propto n^{3/2} \propto (jR)^{18/25}$.


\section{Data availability}
The error correction simulation data shown in this work have been deposited in Ref.~\cite{Pecorari2024}.

\section{Author Contributions}
L.P., G.K.B. and G.P. developed the code modeling and implementation scheme. L.P. performed the error correction simulations. S.J. and G.P. conceptualized the long-range gate design. G.K.B. and G.P. supervised the work. All authors discussed the results and contributed to writing the manuscript.

\section{Competing Interests}
G.P. is co-founder and shareholder of QPerfect. The remaining authors declare no competing interests.

\textcolor{white}{\lipsum[1-2]}
\vspace{15cm}

\appendix
\section{APPENDICES}
\subsection{Appendix A: Details on classical seeds choice}
In this appendix we provide additional details on the quantum code construction discussed in the main text. 

As an example, suppose we want to build a quantum hypergraph product LDPC code with only next-to-next-to-nearest neighbor connectivity. Additionally, we require the stabilizer weight to be low for the ease of near-term implementation with neutral atoms. In fact, a higher number of gates requires a higher number of pulse sequences per atom, which increases the control complexity of the protocol and may ultimately heat up the system increasing the error probability (e.g. atom loss).

Next-to-next-to-nearest neighbor connectivity implies a degree-$3$ seed polynomial $1+x+x^2+x^3\in\mathbb{F}_2[x]/(x^n-1)$ for some code length $n$, while low stabilizer weight imposes $1+x+x^3$. The reason why we keep the coefficient of the $x$ term to be non-zero has already been discussed in the main text. 

We choose classical seeds to be equal, because of constrains on the quantum code parameters, which, we recall, are $D=\text{min}\{d_1,d_2,d_1^T,d_2^T\}$ for the distance and $K=k_1k_2+k_1^Tk_2^T$ for the number of logical qubits. From these expressions it follows that combining via HGP $1+x+x^3$ with any other polynomial of degree $k\leq3$ would give rise to a quantum stabilizer code with lower or at most equal number of logical qubits and lower or at most equal code distance. The last statement follows from the observation that, for a given $n$, classical codes with lower $k$ usually have a larger distance $d$. Therefore, equal seeds choice at the same time guarantees maximal distance and maximal encoding rate, along with square layout (La-cross codes), as the number of physical qubits is $N=(n-k)^2+n^2$. 

As a concrete example, consider now the case where $n=7$. The parity-check matrix, $H$, of the seeds then reads
\begin{equation*}
    H=\begin{pmatrix}
        1 & 1 & 0 & 1 & 0 & 0 & 0 \\
        0 & 1 & 1 & 0 & 1 & 0 & 0 \\
        0 & 0 & 1 & 1 & 0 & 1 & 0 \\
        0 & 0 & 0 & 1 & 1 & 0 & 1 \\
        1 & 0 & 0 & 0 & 1 & 1 & 0 \\
        0 & 1 & 0 & 0 & 0 & 1 & 1 \\
        1 & 0 & 1 & 0 & 0 & 0 & 1
    \end{pmatrix},
\end{equation*}
which can be reduced to the full-rank rectangular matrix  
\begin{equation*}
    H=\begin{pmatrix}
        1 & 1 & 0 & 1 & 0 & 0 & 0 \\
        0 & 1 & 1 & 0 & 1 & 0 & 0 \\
        0 & 0 & 1 & 1 & 0 & 1 & 0 \\
        0 & 0 & 0 & 1 & 1 & 0 & 1
    \end{pmatrix}.
\end{equation*}
The resulting La-cross LDPC codes are $[[98,18,4]]$ and $[[65,9,4]]$, respectively, with quantum parity-check matrix given as discussed in the main text. Here we have used that for equal seed codes the number of physical and logical qubits read $N=2n^2$ and $K=2k^2$ in the first case and $N=(n-k)^2+n^2$ and $K=k^2$ in the second one. From that it can be inferred that the shape of the seed parity-check matrix also determines the boundary conditions of the resulting quantum code. When $H$ is square and circulant, classical checks connect opposite endpoints of the length-$n$ classical code and give rise to a quantum code with stabilizers connecting opposite array boundaries, i.e. with periodic boundary conditions. On the contrary, rectangular parity-check matrices in $\mathbb{F}_2^{(n-k)\times n}$ give rise to a quantum code with stabilizers stretching up to the array extent, i.e. with open boundary conditions. 

\subsection{Appendix B: Gate ordering and hook errors}
In this appendix we show that the chosen gate ordering for performing stabilizer measurements is unique and robust against hook errors.

The choice of gate ordering in syndrome extraction circuits has to obey two main constraints: gate parallelism and error suppression. We here assume the best parallelism possible. Namely, we assume we can measure the four-body plaquette of $X$ and $Z$ stabilizers simultaneously as in the standard surface code \cite{Fowler_2012}. We then perform the four long-range gates sequentially, for a total of four additional circuit steps. This, along with state preparation and measurement, results in the depth-$10$ circuit shown in Fig.~\ref{fig:intro}(b). The requirement that errors should not spread among neighboring stabilizers during the syndrome measurement cycle then uniquely determines the chosen gate ordering, up to long-range gate interchange for $X$ and $Z$ stabilizers.     

We now demonstrate that such syndrome extraction circuit does not suffer from \textit{hook errors} \cite{Dennis_2002}, namely errors occurring on single data qubits throughout the stabilizer measurement cycle and spreading to more than one data qubit at the end of the cycle triggering a logical error. These faults are particularly harmful as they may effectively decrease the distance and consequently degrade the overall quantum error correction performance of the code. One way to avoid hook errors is to carefully choose the gate ordering \cite{Tomita_2014} in such a way that errors affecting single data qubits result in error chains that do not align with the logical operators. 

\begin{figure}[ht]
    \centering
    \includegraphics[width=0.49\textwidth]{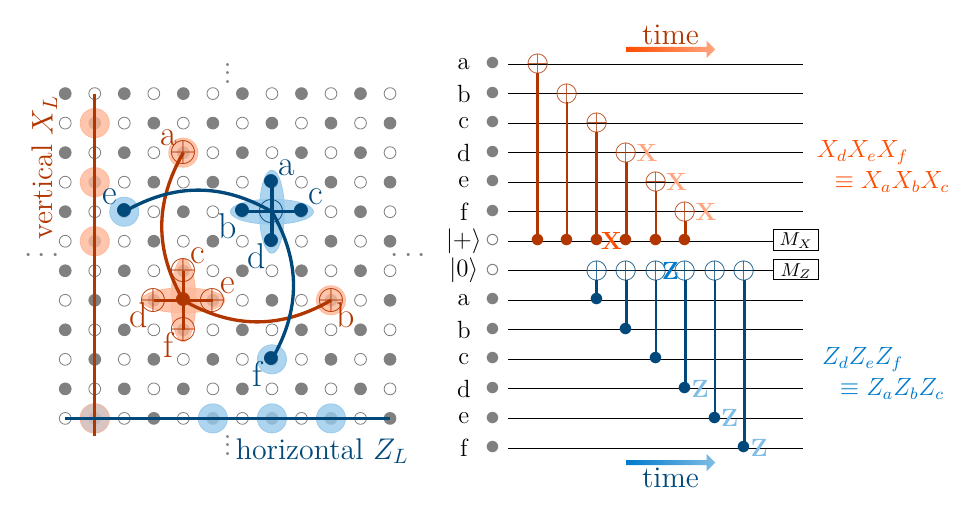}
    \caption{\textbf{Robustness against hook errors.} Left: array patch of $k=3$-LDPC code with two stabilizers and two logical operators, $X_L$ and $Z_L$, of $X$- and $Z$-type. When open boundary conditions are enforced, all $X$-type logical operators are vertical and all $Z$-type ones are horizontal. This protects the chosen syndrome extraction scheme from hook errors. Right: Syndrome extraction circuit for $X$ (red) and $Z$ (blue) stabilizers. An error occurring on the ancilla qubit after three CNOT gates does not trigger a logical error as the corresponding three-qubit error chain does not align with the logical operators modulo stabilizer multiplication.}
    \label{fig:hook}
\end{figure}

In the present case, the above gate ordering ensures that no three-qubit error chains can form, therefore only two-qubit error chains aligned with the logical operators should be considered. 
Let us analyze how errors affecting the ancilla qubit at different time steps propagate to the six data qubits. Let $X_\alpha$ ($Z_\alpha$) with $\alpha\in\{a,b,c,d,e,f\}$ be a resulting single-qubit error affecting the $\alpha$th data qubit as depicted in Fig.~\ref{fig:hook} (right panel) and denote with ``$\equiv$" the equivalence up to stabilizer multiplication. We have
\begin{align*}
    X_aX_bX_cX_dX_eX_f\:(Z_aZ_bZ_cZ_dZ_eZ_f) &\equiv \mathbb{I} \\
    \mathbb{I}_aX_bX_cX_dX_eX_f\:(\mathbb{I}_aZ_bZ_cZ_dZ_eZ_f) &\equiv X_a\;(Z_a) \\ 
    \mathbb{I}_a\mathbb{I}_bX_cX_dX_eX_f\:(\mathbb{I}_a\mathbb{I}_bZ_cZ_dZ_eZ_f) &\equiv X_aX_b\;(Z_aZ_b) \\
    \mathbb{I}_a\mathbb{I}_b\mathbb{I}_c\mathbb{I}_dX_eX_f\:(\mathbb{I}_a\mathbb{I}_b\mathbb{I}_c\mathbb{I}_dZ_eZ_f) &\equiv X_eX_f\;(Z_eZ_f),
\end{align*}
where both $X_aX_b\;(Z_aZ_b)$ and $X_eX_f\;(Z_eZ_f)$ are not aligned with the logical operators. The possibly most dangerous scenario is realized when a $X$ ($Z$) error occurs on a $X$ ($Z$) stabilizer halfway through the stabilizer measurement cycle. Such an error spreads into three data qubit errors at the end of the cycle, two of which being aligned with the horizontal (vertical) $Z_L$ ($X_L$) logical operators. However, we find that this is not an issue as far as open boundary conditions are enforced. In fact, up to stabilizer multiplication, one can always transform such horizontally (vertically) aligned error chains into vertical (horizontal) error chains which do not trigger any logical error, namely
\begin{equation*}
    \mathbb{I}_a\mathbb{I}_b\mathbb{I}_cX_dX_eX_f\:(\mathbb{I}_a\mathbb{I}_b\mathbb{I}_cZ_dZ_eZ_f) \equiv X_aX_bX_c\;(Z_aZ_bZ_c).
\end{equation*}
In fact, logical $X$ ($Z$) operators only enjoy translational invariance under horizontal (vertical) shifts of $2$ lattice sites and not, e.g., under single diagonal shifts on the lattice. This argument ensures that the above error chains never overlap with the logical operators and proves the robustness of our syndrome measurement scheme against hook errors.

On the other hand, when periodic boundary conditions are naively enforced and twice the logical operators are present -- both vertical and horizontal --, the single ancilla stabilizer measurement scheme described here is prone to hook errors. 

\subsection{Appendix C: Comparison of La-cross codes with some quantum LDPC literature}
We here address the issue of comparing quantum code parameters and error correction performance of our La-cross codes with other quantum LDPC codes from the literature. Rather than an exhaustive review, we take into account the main works with a strong near-term implementation focus, namely Refs.~\cite{Bravyi2024,xu2023constantoverhead,poole2024architecturefastimplementationqldpc,Tremblay_2022,Old_2024}. We list these references in Tab.~\ref{tab:literature} and briefly discuss their code parameters, connectivity requirements and geometrical layout. Clearly, the same codes may allow for different implementation schemes, hence may have different connectivity and geometry requirements. We do not discuss error
suppression capabilities, threshold and improvement over the surface code, as these properties intrinsically depend on noise model and metric of comparison. Overall, La-cross quantum LDPC codes offer comparable or better performance and the fastest QEC cycle times thanks to the static implementation protocol, as discussed in the main text.
\begin{table*}[!hb]
\centering
\begin{tabular}{cccc}
\toprule\midrule
quantum LDPC code & Code parameters & Connectivity & Geometry \\
\midrule
Bivariate bicycle codes \cite{bravyi1998quantum,poole2024architecturefastimplementationqldpc} & \makecell{High rate, high $D^*$ \\ {\small $^*$to be searched numerically}}  & long-range & \makecell{biplanar or 2D planar via folding}  \\
Quantum expander codes \cite{Tremblay_2022,xu2023constantoverhead} & Constant  rate & any-to-any & \makecell{reconfigurable 2D planar \\ or stacked planar}\\
Lift-connected surface codes \cite{Old_2024} & \makecell{High rate, $D=\mathcal{O}(N)^*$ \\ {\footnotesize $^*$up to a max. size}}  & 3D local or 2D any-to-any & 3D or reconfigurable 2D planar \\
La-cross codes & \makecell{High rate, $D=\mathcal{O}(\sqrt{N})$}  & long-range & 2D planar \\
\bottomrule
\end{tabular}
\caption{Summary of some of the main near-term-implementation-focused quantum LDPC codes and the here-discussed La-cross codes in terms of code parameters, connectivity requirements and geometry. Same codes may allow for different implementation schemes with different connectivity and geometry requirements.}
\label{tab:literature}
\end{table*}

\end{document}